\documentstyle[12pt]{article}

\begin{document}

\def\fnote#1#2{\begingroup\def\thefootnote{#1}\footnote{#2}
\addtocounter{footnote}{-1}\endgroup}

\hfill{UTTG-29-97}

\vspace{24pt}

\begin{center}
{\large{\bf  Non-Renormalization Theorems in
Non-Renormalizable  Theories}}

\vspace{36pt}

Steven Weinberg\fnote{*}{Research supported in part by the
Robert A. Welch
 Foundation and NSF Grant PHY 9511632.  E-mail address:
weinberg@physics..utexas.edu}

\vspace{4pt}
Theory Group, Department of Physics, University of Texas\\
Austin, TX, 78712\\

\vspace{30pt}
{\bf Abstract}

\end{center}

\begin{minipage}{4.75in}

A perturbative non-renormalization theorem is presented that
applies to general supersymmetric theories, including
non-renormalizable theories in which the $\int d^2\theta$
integrand is an arbitrary gauge-invariant function
$F(\Phi,W)$ of the
chiral superfields $\Phi$ and gauge field-strength
superfields $W$, and the $\int
d^4\theta$-integrand is restricted only by gauge invariance.
In the  Wilsonian Lagrangian, $F(\Phi,W)$ is unrenormalized
except for the one-loop renormalization of the gauge
coupling parameter, and Fayet--Iliopoulos terms can be
renormalized only by one-loop graphs, which cancel if the
sum of the $U(1)$ charges of the chiral superfields
vanishes.   One consequence of this theorem is that in
non-renormalizable as well as renormalizable theories, in
the absence of
Fayet--Iliopoulos terms
 supersymmetry will be unbroken to all orders if the bare
superpotential has a stationary point.

\end{minipage}

\vfill

\baselineskip=24pt
\pagebreak
\setcounter{page}{1}

The remarkable absence of various radiative corrections in
supersymmetric theories was first shown using supergraph
techniques.$^1$  Later Seiberg introduced a simple and
powerful new approach to this problem,$^2$ and used it to
prove non-renormalization theorems in various special
cases.  This paper will use a generalized version of the
Seiberg approach to give a proof of the
perturbative non-renormalization theorems, that applies  not
only to the usual renormalizable theories but also to
general
non-renormalizable theories.  These have a much richer set
of couplings, involving terms of arbitrary order in the
gauge superfield, that are shown to be unrenormalized.

This is of some importance in model building.  As Witten$^3$
pointed out long ago, it is the perturbative
non-renormalization theorems that, by limiting the radiative
corrections responsible for supersymmetry breaking to
exponentially small
non-perturbative terms, offer a hope that supersymmetry
might solve the hierarchy problem.
But the theory with which we have deal below the Planck or
grand-unification scales is surely an effective quantum
field theory, which contains non-renormalizable as well as
renormalizable terms.  Thus, in relying on supersymmetry to
solve the hierarchy problem, we had better make sure that
the non-renormalization theorems apply to non-renormalizable
as well as to renormalizable theories.

We consider a general supersymmetric theory involving
left-chiral superfields $\Phi_n$, their right-chiral
adjoints $\Phi^*_n$, the  matrix gauge superfield $V$, and
their derivatives.  The general supersymmetric action has a
Lagrangian density of the form
\begin{eqnarray}
{\cal L}&=&\int d^2\theta_L\,d^2\theta_R\; \Bigg[
\Big(\Phi^\dagger e^{-
V}\Phi\Big)+G\Big(\Phi,\Phi^\dagger,V,{\cal
D}\cdots\Big)\Bigg] \nonumber\\&&+2\,{\rm Re}\,\int d^2
\theta_L\;\Bigg[\frac{\tau}{8\pi
i}\sum_{\alpha\beta}\epsilon_{\alpha\beta}{\rm Tr}\,
W_\alpha
W_\beta+
F\Big(\Phi,W\Big)\Bigg]\;,
\end{eqnarray}
where $G$ and $F$ are general gauge-invariant functions of
the arguments shown;  `${\cal D}\cdots$' denotes a
dependence of $G$ on superderivatives (or spacetime
derivatives) of the other arguments; $W_\alpha$ is the usual
gauge-covariant matrix left-chiral gauge superfield formed
from $V$; $\alpha$ and $\beta$ are two-component spinor
indices with $\epsilon_{\alpha\beta}$ antisymmetric; and
$\tau$ is the usual complex gauge coupling parameter
\begin{equation}
\tau=\frac{4\pi i}{g^2}+\frac{\theta}{2\pi}\;.
\end{equation}
The function  $F\Big(\Phi,W\Big)$ must be holomorphic in the
left-chiral superfields $\Phi_n$ and $W$,
and may not depend on their superderivatives or spacetime
derivatives because, as well known, any term in
$F\Big(\Phi,W\Big)$ that did depend on these derivatives
could be replaced with a gauge-invariant contribution to
$G$.  The terms $(\Phi^\dagger e^{-V}\Phi)$ and $(\tau/8\pi
i){\rm Tr}\,(W^{\rm T}\epsilon W)$ in Eq.~(1) could have
been included
in $G$ and $F$, respectively; they are displayed here to
identify the zeroth-order kinematic terms that serve as a
starting point for perturbation theory.

The first perturbative non-renormalization theorem to be
proved
here states that the `Wilsonian' effective Lagrangian
density ${\cal L}_\lambda$ (which, with an ultraviolet
cut-off $\lambda$,  yields the same results as the original
Lagrangian density (1)) takes the form
\begin{eqnarray}
{\cal L}_\lambda&=&\int
d^2\theta_L\,d^2\theta_R\;\Bigg[\Big(\Phi^\dagger e^{-
V}\Phi\Big)+ G_\lambda\Big(\Phi,\Phi^\dagger,V,{\cal
D}\cdots\Big) \nonumber\\&&+2\,{\rm Re}\,\int d^2
\theta_L\;\Bigg[\frac{\tau_\lambda}{8\pi
i}\sum_{\alpha\beta}\epsilon_{\alpha\beta}{\rm Tr}\,
W_\alpha
W_\beta+
F\Big(\Phi,W\Big)\Bigg]\;,
\end{eqnarray}
where $G_\lambda$ is some new function of the displayed
variables, and $\tau_\lambda$ is a one-loop renormalized
coupling parameter.  For instance, for a simple gauge group
we have
\begin{equation}
\tau_\lambda=\tau+i\,\left(\frac{3C_1-
C_2}{2\pi}\right)\,\ln(\lambda/\Lambda)\;.
\end{equation}
Here  $\Lambda$ is an integration constant, and $C_1$ and
$C_2$ are the Casimir constants of the gauge and
left-chiral superfields, defined by
\begin{equation}
\sum_{CD}C_{ACD}\,C_{BCD}=C_1\,\delta_{AB}\;,~~~~~~~{\rm
Tr}\,\{t_At_B\}=C_2\,\delta_{AB}\;,
\end{equation}
where $C_{ABC}$ are the structure constants, and $t_A$ are
the matrices representing the gauge algebra on the
left-chiral superfields.    Not only is the the
superpotential
$F(\Phi,0)$ not renormalized --- {\em
the whole $W$-dependent integrand of the $\int d^2\theta_L$
integral is not renormalized, except for a one-loop
renormalization of the gauge coupling constant.}

Here is the proof.  Assuming that the cut-off respects
supersymmetry and gauge invariance,$^4$ these symmetries
require the Wilsonian effective Lagrangian to take the same
general form as Eq.~(1):
\begin{eqnarray}
{\cal L}_\lambda&=&\int d^2\theta_L\,d^2\theta_R\; \Bigg[
\Big(\Phi^\dagger e^{-
V}\Phi\Big)+G_\lambda\Big(\Phi,\Phi^\dagger,V,{\cal
D}\cdots\Big)\Bigg] \nonumber\\&&+2\,{\rm Re}\,\int d^2
\theta_L\;\Bigg[\frac{\tau}{8\pi
i}\sum_{\alpha\beta}\epsilon_{\alpha\beta}{\rm Tr}\,
W_\alpha
W_\beta+
F_\lambda\Big(\Phi,W\Big)\Bigg]\;,
\end{eqnarray}
where $G_\lambda$ and $F_\lambda$ are again general
gauge-invariant functions of the arguments shown, with
$G_\lambda$ Hermitian.  (Since the functions $F_\lambda$ and
$G_\lambda$ are not yet otherwise restricted, there is no
loss of
generality in extracting the terms shown explicitly in
Eq.~(6) from them.)  Following Seiberg,$^2$ we regard $\tau$
and the coefficients $f_r$ of the various terms in
$F(\Phi,W)$ as independent external left-chiral superfields
that  happen to have constant scalar components and no
spinor or auxiliary components, and that should also appear
among the arguments of $F_\lambda$ and (along with their
adjoints) in $G_\lambda$.

To deal with the non-renormalizable part of the $\int
d^4\theta$ integral, we now also regard the coefficients of
the various terms in the real function
$G\Big(\Phi,\Phi^\dagger,V,{\cal D}\cdots\Big)$ as {\em
real} external superfields, which also have only constant
scalar components.  It is tempting to say that because these
real superfields are non-chiral, they cannot appear in the
integrand of the $\int d^2\theta_L$ integral in ${\cal
L}_\lambda$, so that $F_\lambda$ may be analyzed as if $G$
were not present.  This would be too hasty, because any real
superfield $P$ with
a positive scalar component may
be expressed in terms of a left-chiral superfield $Z$, as
\begin{equation}
P=Z^*Z\exp(V_P)\;,
\end{equation}
where $V_P$ has the
form of a $U(1)$ gauge superfield in any fixed gauge. (Note
that  $\ln P\rightarrow \ln P-\ln Z-\ln Z^*$
is a
generalized gauge transformation.)
Eq.~(7) is
invariant under a phase transformation $Z\rightarrow
Ze^{i\alpha}$, which if unbroken would prevent the
left-chiral superfield $Z$
from appearing in $F_\lambda$.  This symmetry is
actually
violated by non-perturbative effects.  For instance, if we
modify the usual renormalizable kinematic term for a
multiplet of left-chiral superfields to read $\int
d^2\theta_L d^2\theta_R Z^*Z(\Phi^*e^{-V}\Phi)$, then since
this depends only on $Z\Phi$, the transformation
$Z\rightarrow Ze^{i\alpha}$, $\Phi\rightarrow \Phi$ has the
same anomaly as
the transformation $\Phi\rightarrow \Phi e^{i\alpha}$,
$Z\rightarrow Z$.  This
anomaly leads to a breakdown of this symmetry, which allows
$Z$ to appear in
non-perturbative corrections to the superpotential.  Indeed,
if it were not for this breakdown of the symmetry under
$Z\rightarrow Ze^{i\alpha}$, there could be no
non-perturbative corrections to the superpotential, because
the kinematic term is invariant under a non-anomalous
transformation $Z\rightarrow Ze^{i\beta}$,  $\Phi\rightarrow
\Phi e^{-i\beta}$, which would prevent the generation  of a
non-perturbative term in the superpotential that depends on
$\Phi$ but not $Z$.  Here we are considering only
perturbation theory, so $Z$ cannot appear in $F_\lambda$,
and by the same reasoning  neither can any of the real
superfields that appear as coefficients of the terms in $G$.
Thus $G$ can have no effect on $F_\lambda$.

With $G$ ignored, the perturbation theory based on the
Lagrangian density (1) has two symmetries that restrict the
dependence of $F_\lambda$ on $\tau$ and on the parameters
$f_r$ in $F(\Phi,W)$.  The first symmetry is conservation of
an $R$ quantum number: $\theta_L$ and $\theta_R$ have $R=+1$
and $R=-1$; $\tau$ and the $\Phi_m$ and $V$ have $R=0$; and
the  coefficients  $f_r$ of all terms in $F(\Phi,W)$ with
$r$ factors of $W_\alpha$ have $R=2-r$.  (Since $W_\alpha$
involves two superderivatives of $V$ with respect to
$\theta_R$ and one with respect to $\theta_L$, it has
$R=+1$.  Also, in accordance with the usual rules for
integration over Grassman parameters, integration of a
function over $\theta_L$ {\em lowers} its $R$ value by two
units.)    This symmetry requires the function
$F_\lambda(\Phi,W,f,\tau)$ to have $R=2$, like $F(\Phi,W)$.
The other symmetry is invariance under translation of $\tau$
by an arbitrary real number $\xi$:
\begin{equation}
\tau\rightarrow \tau+\xi\;,
\end{equation}
which leaves the action invariant because ${\rm Im}\,\int
d^2\theta_L \sum_{\alpha\beta}\epsilon_{\alpha\beta}\,{\rm
Tr}\,W_\alpha
W_\beta$ is a spacetime derivative.   This tells us that
$F_\lambda$ is independent of $\tau$, except for a possible
term proportional to the $WW$ term in $F$:
\begin{equation}
F_\lambda(\Phi,W,f,\tau)=c_\lambda\tau
\sum_{\alpha\beta}\epsilon_{\alpha\beta}{\rm Tr}\, W_\alpha
W_\beta
+H_\lambda(\Phi,W,f)\;,
\end{equation}
with $c_\lambda$ a real function of $\lambda$.

To use this information,  let us consider how many powers of
$\tau$ are contributed to $F_\lambda$ by each graph.
Suppose a graph has $E_V$ external left-handed gaugino lines
and any number of external $\Phi$-lines; $I_V$ internal
$V$-lines and any number of internal $\Phi$-lines;
${\cal A}_m$ pure gauge vertices with $m\geq 3$ $V$-lines,
arising from
the $WW$ term in Eq.~(1); ${\cal B}_{mr}$ vertices with
$m\geq r$ $V$-lines and any number of $\Phi$-lines, arising
from the terms in $F(\Phi,W)$ with $r$ factors of $W$; and
${\cal C}_m$ vertices with two $\Phi$-lines and $m\geq 1$
$V$-lines, arising from the $\Phi^\dagger e^{-V}\Phi$ term
in Eq.~(1).
(By `$\Phi$-lines' and `$V$-lines' are meant lines of the
component fields of left-chiral or gauge superfields,
respectively; these are ordinary Feynman graphs, not
supergraphs.)  These numbers are related by
\begin{equation}
2I_V+E_V=\sum_{m\geq 3} m{\cal A}_m+\sum_r\sum_{m\geq
r}m{\cal B}_{mr}+\sum_{m\geq 1}m{\cal C}_m\;.
\end{equation}
Also, since we have specified that all external $V$ lines
are for gauginos, this graph can only contribute to a term
in ${\cal L}_\lambda$ with $E_V$ factors of $W$, so it must
be proportional to a product of $f_r$ factors with total
$R$-value $2-E_V$, and so
\begin{equation}
\sum_r\sum_{m\geq r}(2-r){\cal B}_{mr}=2-E_V\;.
\end{equation}
Using this to eliminate $E_V$ in Eq.~(10), the number of
factors of $\tau$ contributed by such a graph is then
\begin{eqnarray}
N_\tau&=&\sum_{m\geq 3}{\cal A}_m-I_V \nonumber\\&=&1-
\frac{1}{2}\left[\sum_{m\geq 3} (m-2){\cal
A}_m+\sum_r\sum_{m\geq r}(2-r+m){\cal B}_{mr}+\sum_{m\geq
1}m\,{\cal C}_m\right]\;.~~~~~~
\end{eqnarray}
Each of the ${\cal A}$s, ${\cal B}$s, and ${\cal C}$s in the
square brackets in Eq.~(12) has a positive-definite
coefficient, so there is a limit to the number of vertices
of each type that can contribute to the $\tau$-independent
function $H_\lambda$.  To have $N_\tau=0$, we can have
${\cal A}_3=2$ and all other ${\cal A}$s, ${\cal B}$s, and
${\cal C}$s zero, or ${\cal A}_4=1$ and all other ${\cal
A}$s, ${\cal B}$s, and
${\cal C}$s zero, which give the one-gauge-loop
contributions proportional to $C_1$ in Eq.~(4); or ${\cal
B}_{mr}=1$ for some $r$ and $m=r$, and all other ${\cal
A}$s, ${\cal B}$s, and ${\cal C}$s zero, which add up to the
one-vertex tree contribution $F(\Phi,W)$ to the function
$H_\lambda(\Phi,W)$ in Eq.~(9); or ${\cal C}_1=2$ and all
other ${\cal A}$s, ${\cal B}$s, and ${\cal C}$s zero, or
${\cal C}_2=2$ and all other ${\cal
A}$s, ${\cal B}$s, and ${\cal C}$s zero, which
give the one-$\Phi$-loop contributions proportional to $C_2$
in Eq.~(4).  (Graphs with ${\cal A}_3={\cal C}_1=1$ and all
other ${\cal A}$s, ${\cal B}$s, and ${\cal C}$s zero are
one-particle-reducible, and therefore do not contribute to
${\cal L}_\lambda$.)  Finally, Eq.~(12) shows that there are
no graphs at all with $N_\tau=1$, so the constant
$c_\lambda$ in Eq.~(9) vanishes, completing the proof of the
non-renormalization theorem (3).

In theories where the gauge group has a $U(1)$ factor, there
is also one term in $G_\lambda$ which is  subject to a
non-renormalization theorem.  As pointed out by Fayet and
Iliopoulos,$^5$ although a $U(1)$ gauge superfield $V_1$ is
not gauge invariant, $\int d^4x\int d^4\theta\,V_1$ is gauge
invariant as well as supersymmetric, so we can include a
term $\xi V_1$ in $G$.  By detailed calculation, Fischler et
al.$^6$ showed that the constant $\xi$ receives corrections
for renormalizable theories  only from one-loop diagrams,
and that these corrections vanish if the sum of the $U(1)$
charges of the left-chiral superfields vanish.    Using the
Seiberg trick of regarding coupling
parameters as the scalar components of external superfields,
it is easy to give a very simple proof$^7$ of this result,
which
applies also in non-renormalizable theories, and even
non-perturbatively.
The point is, that a term $\int d^4x\int d^4\theta\,S\,V_1$
in $G_\lambda$ is {\em not} gauge-invariant if $S$ depends
in a non-trivial way on any superfields, including the
external superfields $\tau$ or $f_r$ or those appearing as
coefficients of
the non-renormalizable terms in $G$.  There is just one
graph that can make a correction to $\xi$ that is
independent of all coupling constants: it is the one-loop
tadpole graph, in which an external $V_1$
line interacts with left-chiral superfields through the
term $(\Phi^\dagger \exp(-V)\Phi)$ in Eq.~(1).  This graph
vanishes if the sum of the $U(1)$ charges of the left-chiral
superfields vanish.  This condition is necessary (unless the
$U(1)$ gauge symmetry is spontaneously broken) to avoid
gravitational anomalies$^8$ that would violate the
conservation of the $U(1)$ current.

What good are the non-renormalization theorems, when so
little  is known about
the structure of the function
$G_\lambda(\Phi,\Phi^\dagger,V,{\cal D}\cdots)$?
Fortunately, in the absence of Fayet--Iliopoulos terms, it
turns out that only the bare superpotential matters in
deciding if supersymmetry is spontaneously broken:  if the
superpotential $F(\Phi,0)$
allows solutions of the equations $\partial
F(\Phi,0)/\partial \Phi_n=0$ then
supersymmetry is not broken in any finite order of
perturbation theory.

To test this, we must examine
Lorentz-invariant field configurations, in which the
$\Phi_n$
have only constant scalar components $\phi_n$ and constant
auxiliary auxiliary components ${\cal F}_n$, while
(in Wess--Zumino gauge) the coefficients $V_A$ of the gauge
generators $t_A$ in the matrix gauge superfield $V$ have
only  auxiliary components
$D_A$.
Supersymmetry is unbroken if there are values of $\phi_n$
for which ${\cal L}_\lambda$ has no terms of first order in
${\cal F}_n$ or $D_A$, in which case there is sure to be an
equilibrium solution with ${\cal F}_n=D_A=0$.  In the
absence of Fayet--Iliopoulos terms, this requires that for
all $A$
\begin{equation}
\sum_{nm}\frac{\partial
 K_\lambda(\phi,\phi^*)}
{\partial\phi^*_n}
(t_A)_{mn}\phi^*_m=0\;,
\end{equation}
and for all $n$
\begin{equation}
\frac{\partial F(\phi,0)}{\partial\phi_n}=0\;,
\end{equation}
where the effective Kahler potential
$K_\lambda(\phi,\phi^*)$ is
\begin{equation}
K(\phi,\phi^*)=\Big(\phi^\dagger\phi\Big)+
G_\lambda(\phi,\phi^*,0,0\cdots)
\end{equation}
with $G_\lambda(\phi,\phi^*,0,0\cdots)
$ obtained from
$G_\lambda$ by setting the gauge superfield and all
superderivatives equal
to zero.  (With superderivatives required to vanish by
Lorentz invariance, the only dependence of $G_\lambda$ on
$V$ is a factor $\exp(-V)$ following every factor
$\Phi^\dagger$.)  If there is any solution $\phi^{(0)}$ of
Eq.~(14), then
the gauge symmetry tells us that there is a continuum of
such solutions, with $\phi_n$ replaced with
\begin{equation}
\phi_n(\xi)=\Big[\exp(i\sum_A
t_A\xi_A)\Big]_{nm}\phi^{(0)}_m
\end{equation}
where (since $F$ depends only on $\phi$, not $\phi^*$), the
$\xi_A$ are an arbitrary set of {\em complex} parameters.
If $K_\lambda(\phi,\phi^*)$ has a stationary point anywhere
on the surface $\phi=\phi(\xi)$, then at that point
\begin{equation}
0=\sum_{nmA}\frac{\partial
K_\lambda(\phi,\phi^*)}{\partial
\phi_n}(t_A)_{nm}\phi_m\,\delta
\xi_A
-\sum_{nmA}\frac{\partial
K_\lambda(\phi,\phi^*)}{\partial
\phi^*_n}(t_A)_{mn}\phi^*_m\,\delta
\xi^*_A
\;.
\end{equation}
Since this must be satisfied for all infinitesimal {\em
complex}
$\delta\xi_A$,
the coefficients of both $\delta\xi_A$ and $\delta\xi_A^*$
must both vanish, and therefore Eq.~(13) as well as Eq.~(14)
is satisfied at this point.  Thus the existence of a
stationary point of $K_\lambda(\phi,\phi^*)$ on the surface
$\phi=\phi(\xi)$ would imply
that supersymmetry is unbroken
to all orders of perturbation theory.  The zeroth-order
Kahler potential $(\phi^\dagger\phi)$ is bounded below and
goes to infinity as $\phi\rightarrow \infty$, so it
certainly has a minimum on the surface $\phi=\phi(\xi)$,
where of course it is stationary.  At this minimum there are
flat directions: ordinary  global gauge transformations
$\delta \phi=i\sum_A\delta\xi_A t_A\phi$ with $\xi_A$ real.
But these are also flat directions for the perturbation
$G_\lambda(\phi,\phi^*,0,0)$.  Thus there is still a local
minimum
of $K_\lambda$ on the surface $\phi=\phi(\xi)$ for  any
perturbation
$G_\lambda(\phi,\phi^*,0,0)$ in at least a finite range, and
thus to all orders in whatever couplings appear in
$G_\lambda(\phi,\phi^*,0,0)$.  We see that in such a theory
supersymmetry can only be broken by non-perturbative
effects, which can naturally lead to a solution of the
hierarchy problem.

I am glad to acknowledge  helpful conversations  with J.
Distler, W. Fischler, V. Kaplunovsky, and N. Seiberg.
\pagebreak

\noindent
{\bf References}

\begin{enumerate}

\item M. T. Grisaru, W. Siegel, and M. Ro\v{c}ek,
{\em Nucl. Phys.} {\bf B159}, 429 (1979).

\item N. Seiberg, {\em Phys. Lett.} {\bf B318}, 469 (1993).

\item E. Witten, {\em Nucl. Phys.} {\bf B185}, 313 (1981).

\item T. Hayashi, Y. Ohshima, K. Okuyama, and H. Suzuki,
Ibaraki preprint IU-MSTP/27, hep-th/9801062, and earlier
references cited therein.

\item P. Fayet and J. Iliopoulos, {\em Phys. Lett.} {\bf
51B}, 461 (1974).

\item W. Fischler, H. P. Nilles, J. Polchinski, S. Raby, and
L. Susskind, {\em Phys. Rev. Lett.} {\bf 47}, 757 (1981).

\item After this paper was submitted for publication I found
that the same proof of the non-renormalization of the
Fayet--Iliopoulos term has been given by M. Dine, in {\em
Fields, Strings, and Duality: TASI 96}, eds. C. Efthimiou
and B. Greene (World Scientific, Singapore, (1997).

\item R. Delbourgo and A. Salam, {\em Phys. Lett.} {\bf
40B}, 381 (1972); T. Eguchi and P. Freund, {\em Phys. Rev.
Lett.} {\bf 37}, 1251 (1976).  For a review and other
references, see T. Eguchi, P. B. Gilkey, and A. J. Hanson,
{\em Physics Reports} {\bf 66}, 213 (1980).
\end{enumerate}
\end{document}